\documentclass[aps,pra,twocolumn,groupedaddress,nofootinbib]{revtex4-1}
\usepackage{graphicx}  % needed for figures
\usepackage{amsmath}
\newcommand{\ket}[1]{|#1\rangle}

\begin{document}

% Use the \preprint command to place your local institutional report
% number in the upper righthand corner of the title page in preprint mode.
% Multiple \preprint commands are allowed.
% Use the 'preprintnumbers' class option to override journal defaults
% to display numbers if necessary
%\preprint{}

%Title of paper
%\title{Differential clock comparisons with phase-locked local oscillators}
%\title{Surpassing the laser noise limits in optical clock comparisons}
\title{Probing beyond the laser coherence time in optical clock comparisons}

% repeat the \author .. \affiliation  etc. as needed
% \email, \thanks, \homepage, \altaffiliation all apply to the current
% author. Explanatory text should go in the []'s, actual e-mail
% address or url should go in the {}'s for \email and \homepage.
% Please use the appropriate macro foreach each type of information

% \affiliation command applies to all authors since the last
% \affiliation command. The \affiliation command should follow the
% other information
% \affiliation can be followed by \email, \homepage, \thanks as well.
\author{David B. Hume}
\email[]{david.hume@nist.gov}
\author{David R. Leibrandt}
%\homepage[]{Your web page}
%\thanks{}
%\altaffiliation{}
\affiliation{National Institute of Standards and Technology, 325 Broadway, Boulder, CO 80305, USA}

%Collaboration name if desired (requires use of superscriptaddress
%option in \documentclass). \noaffiliation is required (may also be
%used with the \author command).
%\collaboration can be followed by \email, \homepage, \thanks as well.
%\collaboration{}
%\noaffiliation

\date{\today}

\begin{abstract}
We develop differential measurement protocols that circumvent the laser noise limit in the stability of optical clock comparisons by synchronous probing of two clocks using phase-locked local oscillators. This allows for probe times longer than the laser coherence time, avoids the Dick effect, and supports Heisenberg-limited measurement precision. We present protocols for such frequency comparisons and develop numerical simulations of the protocols with realistic noise sources.  These methods provide a route to reduce frequency ratio measurement durations by more than an order of magnitude.
\end{abstract}

%\begin{abstract}
%We develop protocols that circumvent the laser noise limit in optical clock comparisons by synchronous probing of two clocks using phase-locked local oscillators. This allows for probe durations longer than the laser coherence time, avoids the Dick effect, and supports Heisenberg-limited scaling of measurement precision. We present a model for such frequency comparisons and develop numerical simulations of the protocol with realistic noise sources.  This provides a route to reduce the duration of many frequency ratio measurements by more than an order of magnitude.
%\end{abstract}

% insert suggested PACS numbers in braces on next line
% insert suggested keywords - APS authors don't need to do this
%\keywords{}

%\maketitle must follow title, authors, abstract, \pacs, and \keywords
\maketitle

\section{\label{sec:intro}Introduction}
Optical clock measurements are the most stable measurements of any kind~\cite{Hinkley2013,Nicholson2015,AlMasoudi2015}, driven largely by recent progress in ultrastable  lasers~\cite{Jiang2011, Kessler2012,Hafner2015}. Still, laser frequency noise limits the stability of frequency comparisons well short of the limits imposed by atomic coherence~\cite{Chou2011}, and has so far prevented the use of Heisenberg-limited measurements that realize a quantum enhancement in measurement stability~\cite{Bollinger1996, Huelga1997}.  High-stability optical clock comparisons are critical for the future redefinition of the SI second~\cite{Gill2011,Riehle2015} and provide a key measurement tool for the parameters of fundamental physical theories~\cite{Uzan2003, Uzan2011, Derevianko2014}, as well as relativistic geodesy with high spatial and temporal resolution~\cite{Chou2010b,Velicogna2005,Chen2014a}.  While there has been a lot of recent progress both towards improving the frequency stability of clock lasers and developing measurement protocols aimed at circumventing clock laser noise using multiple atomic ensembles~\cite{Rosenband2013,Borregaard2013,Kessler2014PRL, Kohlhaas2015}, it is likely that for the foreseeable future optical clock stability will continue to be limited by local oscillator noise.

It is important to recognize, however, that none of the clock applications mentioned above require good absolute (i.e., single) clock stability.  For two clocks operating at the same frequency, it is possible to have better clock comparison stability than absolute clock stability.  For example, clock comparison instability due to the Dick effect~\cite{Dick1987,Santarelli1998,Westergaard2010} can be circumvented by synchronous interrogation of two atomic ensembles with a single local oscillator (LO), which has been demonstrated for microwave~\cite{Bize2000, Biedermann2013, Meunier2014} as well as optical clocks~\cite{Takamoto2011,Takamoto2015}.  A related technique uses a single clock laser to simultaneously probe two clock atoms and derives an error signal from correlations in the transition probabilities between the two~\cite{Chwalla2007,Chou2011}, allowing the probe time to extend beyond the laser coherence time.  Here, we expand these ideas to the more general case of frequency comparisons between clocks operating at different frequencies.   We take advantage of the fact that the relative phase between two local oscillators, even if they are separated by optical frequencies, can, in general, be stabilized more precisely than the absolute phase~\cite{Foreman2007,Nicolodi2014}.  We show that, by using phase-locked LOs and synchronous probing of multiple clocks, optical clock comparisons can operate near the limits imposed by atomic coherence and achieve Heisenberg-limited performance even in the presence of laser noise.

In what follows, we consider the use of this technique in several relevant regimes of optical frequency measurements, distinguished primarily by the number of atoms in each of the two clocks.  We compare the achievable stability in these measurements to what can be achieved in a typical measurement protocol with independent LOs and asynchronous probing, but otherwise identical clock parameters.  First, we introduce the analytic (Sec.~\ref{sec:analytic}) and numerical (Sec.~\ref{sec:numerical}) calculations, focusing on the case when the projection noise of clock 1 is much lower than that of clock 2.  This is relevant, for example, in a frequency comparison between a single-ion clock and a optical lattice clock.  In Sec.~\ref{sec:GHZ}, we extend this discussion to the case where a small number of trapped ions are prepared in an entangled state.  In Sec.~\ref{sec:maximum-likelihood} we further extend our protocol to the case when both clocks have many atoms, as would be true, for example, in a measurement between two optical lattice clocks.  In all cases we find a significant improvement in the measurement stability in the presence of realistic LO noise compared to the usual measurement protocol with independent clocks.

%Differential spectroscopy allows us to unwrap the phase evolution of clock 2, which may fall outside the range $[-\pi/2, \pi/2]$ due to laser frequency noise.

\section{\label{sec:analytic}Analytic estimates of clock stability}

The standard quantum limit (SQL), also known as the projection noise limit, for an atomic clock using Ramsey spectroscopy on $N$ uncorrelated atoms can be written as
\begin{equation}\label{eq-proj-noise-1}
\left(\frac{\Delta\nu}{\nu}\right)^2 = \frac1{(2 \pi\nu)^2NT\tau},
\end{equation}
where $\nu$ is the atomic transition frequency, $T$ is the Ramsey probe duration, and $\tau$ is the total measurement duration~\cite{Itano1993}.  Local oscillator noise constrains clock stability by limiting $T$ to some fraction $\eta$ of the LO coherence time~\cite{Rosenband2012}, which is often much shorter than the atomic coherence time.  If the LO noise is predominantly flicker with a fractional frequency instability $\sigma_L$, we optimize the stability of the atomic clock by choosing ${T = \eta/(\nu\sigma_L)}$.

In a typical frequency ratio measurement two LOs are stabilized independently to two atomic ensembles, and their frequency ratio is measured using a frequency comb.  The clock stability is optimized on clock $j$ by maximizing the probe duration $T_j$ while ensuring that the relative phase between the LO and the atoms, given by
\begin{equation}\label{eq-atom-phase1}
\phi_{j} = 2\pi\int_0^{T_j} \left[\nu_j -  f_j(t) \right] dt = 2\pi \left( \nu_j - \bar{f}_j \right) T_j \ ,
\end{equation}
does not exceed the range $[-\pi/2, \pi/2]$, where $\bar{f}_j$ is the mean frequency of LO $j$ during the probe duration. The measurement variance in the frequency ratio is just the sum of uncorrelated contributions from the two clocks as described by Eq.~(\ref{eq-proj-noise-1}).

\begin{figure}[t!]
    \begin{center}
        \includegraphics{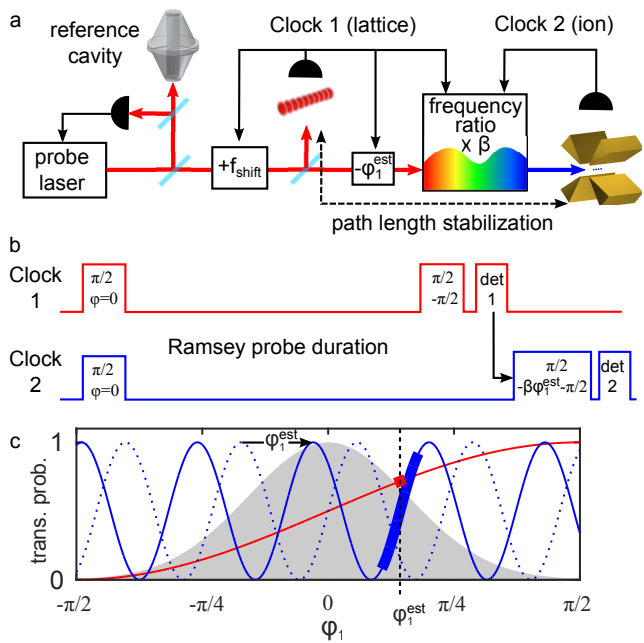}
        \caption{\label{figs-transition-probability}Optical clock comparison with phase-locked LOs. (a) A cavity-stabilized laser simultaneously probes clocks at two different frequencies, which are phase locked via a frequency comb and active path-length stabilization.  (b) Timing diagram of a near-synchronous Ramsey experiment.  The phase $\phi_1^{\rm est}$ measured at clock 1 is used to correct the laser phase before the final pulse of the second clock.  (c) Transition probabilities as a function of $\phi_1$ for clock 1 (red) and clock 2 before and after applying the feed-forward phase (blue dotted and blue solid lines, respectively).  The size of the projection noise for the two clocks is denoted by the thicker lines, and the distribution of laser phase noise is depicted as the gray region.}
    \end{center}
\end{figure}

Now consider the case that clock 1 and clock 2 are probed simultaneously with phase-locked LOs (see Fig.~\ref{figs-transition-probability}), so that their frequencies $f_1(t)$ and $f_2(t)$ are related exactly by a known ratio $\beta = f_2(t)/f_1(t)$, and the noise in the phase measurements is correlated.  The phase evolution of clock 2 during the probe can be written as
\begin{equation}\label{eq-atom-phase2}
\phi_{2} = \frac{\nu_2}{\nu_1} \phi_{1} + 2 \pi T \bar{f}_{1} \epsilon,
\end{equation}
where $\epsilon = \nu_2/\nu_1 - \beta$ is the current error in the frequency ratio measurement.  The first term in Eq.~(\ref{eq-atom-phase2}) correlates the phase measurements on the two clocks and will dominate $\phi_{2}$ when the static ratio $\nu_2/\nu_1$ is sufficiently well known.  In the presence of this correlated noise, information from the measurement of clocks 1 and 2 can be combined to relax the restriction $|\phi_j| < \pi/2$, such that one or both clocks can be operated beyond their laser coherence time.

%Lattice-ion protocol (, which will be an important measurement for verifying the accuracies of both types of clocks in the coming years.)
We illustrate this idea by considering a comparison between a clock with $N_1$ atoms at frequency $\nu_1$ and a single-atom clock ($N_2 = 1$) at frequency $\nu_2>\nu_1$.  This describes, for example, the comparison between an optical lattice clock and a single-ion clock. For a typical (asynchronous) clock comparison, with ${N_1\gg\nu_2/\nu_1}$ and no dead time in either clock,  the measurement noise is dominated by the projection noise of clock 2.  This is limited by the condition ${T_2 = \eta/(\sigma_L\nu_2)}$, such that ${(\Delta\beta/\beta)^2 \simeq \sigma_L/(4\pi^2\eta\nu_2\tau)}$.  With simultaneous probes and phase-locked LOs, the measured value of $\phi_1$ can be used to unwrap the measured value of $\phi_2$ via Eq.~(\ref{eq-atom-phase2}) and extend the clock 2 probe duration to, ${T_2 = T_1 = \eta/(\sigma_L\nu_1)}$, with a corresponding reduction in the measurement variance ${R_{\beta}\equiv(\Delta\beta^{\prime}/\Delta\beta)^2 \simeq \nu_1/\nu_2}$.  One way to do this is illustrated in Fig.~\ref{figs-transition-probability}, where the atom-laser phase difference $\phi_{1}^{\rm est}$ is applied as a feed-forward correction to the laser before the measurement on clock 2.  This measurement of ${\phi_2-(\nu_2/\nu_1)\phi_{1}^{\rm est}}$ is then a differential phase measurement between the two clocks, which is kept in the invertible range ${|\phi_2-(\nu_2/\nu_1)\phi_{1}^{\rm est}| < \pi/2}$.

\begin{figure}
\begin{center}
\includegraphics{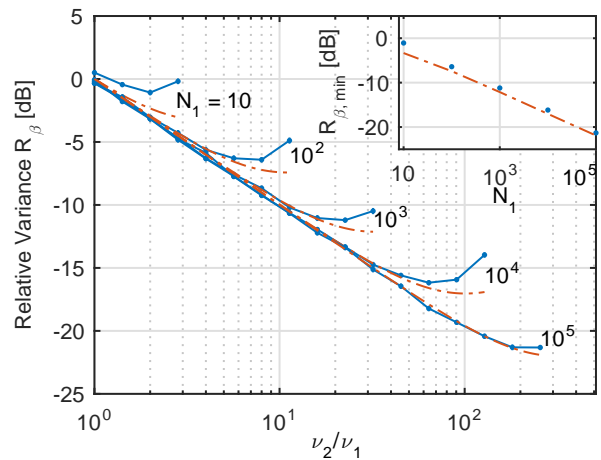}
\caption{\label{figs-noise-vs-ratio}Noise reduction for a frequency ratio measurement of a many-atom clock with a single atom clock.  Precise laser phase measurements on clock 1 allow the unambiguous determination of the clock 2 phase for probe durations longer than the laser coherence limit, giving a reduction of measurement noise compared to the projection noise limit for asynchronous clock comparisons with otherwise identical noise.  Simulation results (points) reproduce the analytical estimates (dash-dotted lines) up to the point that the projection noise for the two clocks is comparable.  Inset: Minimum relative variance $R_{\beta, {\rm min}}$ plotted vs $N_1$, showing that higher atom numbers support greater suppression of the noise.}
\end{center}
\end{figure}

The expected reduction of projection noise in this protocol for different atom numbers $N_1$ has been plotted as the dash-dotted lines in Fig.~\ref{figs-noise-vs-ratio} where we have included the projection noise contributions from both clocks.  In addition to the reduction of projection noise plotted, Dick effect noise is absent for the differential measurement, even in the presence of dead time.  As shown, the available stability improvement using this protocol scales with the frequency ratio, but it must be supported by a sufficiently precise measurement of $\phi_1$, requiring ${\sqrt{N_1} \gg \nu_2/\nu_1}$.  Numerical simulation results, as described below, are plotted along with the analytical estimates in Fig~\ref{figs-noise-vs-ratio}.

\section{\label{sec:numerical}Numerical model}

The arguments outlined above give a conceptual picture of the differential clock comparison protocols that we propose.  The purpose of these protocols is to make optical frequency comparisons immune to the dominant sources of laser noise that limit the comparison stability.  To include, in detail, laser noise with realistic noise spectra we develop here a Monte-Carlo simulation of the protocols that makes use of experimentally demonstrated values for all noise contributions\footnote{Our simulations do not include magnetic field noise, which we expect can be shielded to a negligible level for the clock transitions considered. Alternatively, for some elements, bosonic isotopes with zero first-order magnetic field sensitivity might be used.  We also do not consider averaging of multiple transition frequencies with different magnetic quantum numbers, as is often done to eliminate first-order sensitivity to magnetic fields~\cite{Ludlow2015}.  To the extent that the atoms are well-shielded from magnetic field fluctuations and the overall clock duty-cycle is not affected by switching between magnetic sublevels, this should not affect our results.}, taken from the literature.  In what follows we describe the basic numerical model, and its application to the lattice-ion measurement described in Sec.~\ref{sec:analytic}.   In Secs.~\ref{sec:GHZ} and~\ref{sec:maximum-likelihood} it is adapted to other frequency ratio measurement scenarios.

The laser frequency noise in these simulations is designed to reproduce noise spectra representative of state-of-the-art clock lasers \cite{Jiang2011, Nicholson2012}.  Similarly, differential noise between the two probe lasers is modeled based on published results for active path-length stabilization~\cite{Foreman2007} and coherence transfer through a femtosecond frequency comb~\cite{Nicolodi2014}.  During each clock cycle, both correlated and differential laser frequency noise is generated by filtering pseudorandom white noise in the Fourier domain \cite{Lennon2000}.  The Dick effect in these simulations arises naturally when we introduce dead time to the clock.  Specific values for the parameters of the model are provided in the Appendix (Table~\ref{tab:MCparams}).

The laser frequency for each clock, labeled $j$, can be written as,
\begin{equation}
f_j(t) = \nu_j+n_j(t)+c_j(t),
\end{equation}
where $\nu_j$ is the static atomic resonance frequency, $n_j(t)$ is the laser noise term, and $c_j(t)$ is the frequency correction, which is updated at the end of each clock cycle. Each clock cycle, labeled below with $k$, consists of the clock probe duration followed by dead time required for steps in the experimental sequence such as detection, loading, laser cooling and state preparation.  We have assumed for all of our simulations that the durations of the Ramsey $\pi/2$ pulses are short compared to the Ramsey probe duration $T$.  The time-averaged frequency of the clock $j$ laser during cycle $k$, given by $\bar{f}_{j, k} = \nu_j + \bar{n}_{j, k} + c_{j, k}$, is used to model the atom-laser phase evolution via Eq.~(\ref{eq-atom-phase1}).

Typically, the phase of the second Ramsey pulse is shifted by $-\pi/2$ with respect to the first.  If we include a finite excited-state lifetime $\tau$, the atomic transition probability is given by
\begin{equation}
R(\phi) = \left( 1 + e^{-T/(2 \tau)} \sin \phi \right)/2,
\end{equation}
and its inverse is given by
\begin{equation}
R^{-1}(p) = \arcsin \left[e^{T/(2 \tau)} \left( 2 p - 1 \right)\right],
\end{equation}
with $R^{-1}(p)\in[-\pi/2,\pi/2]$\footnote{In the Monte Carlo model, due to projection noise, it is possible for $|e^{T/(2 \tau)} \left( 2 p - 1 \right)| > 1$, in which case $R^{-1}(p)$ is taken to saturate the bounds given.}.  We estimate the phase of clock $j$ during probe $k$ using the measurement result ${p_{j,k} = \frac{M_{j,k}}{N_j}}$, where $N_j$ is the total number of atoms and $M_{j,k}$, randomly selected from a binomial distribution, is the number of atoms measured to be in spin up.  In some cases, an additional measurement phase $\theta_{j,k}$ is applied and must be accounted for in the phase inversion.  In this case,
\begin{equation}\label{eqn-simple-phase-estimate}
\phi_{j,k}^{est} = R^{-1}(p_{j,k}) - \theta_{j,k}\ .
\end{equation}
For Fig. 2, for example, we have $\theta_{2,k} = -\beta_k\phi_{1,k}^{\rm est}$ from the feed-forward correction to the laser.   By properly accounting for the measurement phases, the phases $\phi_{1, k}^{\rm est}$ and $\phi_{2, k}^{\rm est}$ estimate the real atom-laser phase evolution given in Eq.~\ref{eq-atom-phase1}.

%Here, a non-zero clock 1 measurement and phase feedforward time is included by extending the Ramsey probe time of clock 2 to be longer than that of clock 1

In our protocols, we take advantage of the fact that much of the noise in these estimates is common mode, and we correct the ratio using only the differential component of the atomic phase measurements.  For the $k$th probe, we set $\beta_k$ to be equal to our current best knowledge of the actual atomic transition frequency ratio, which is updated according to
\begin{equation}\label{eq-freq-ratio-update}
\beta_{k+1} = \beta_k - \frac{G_\beta}{2 \pi T \tilde{\nu}_1} ( \phi_{2,k}^{est}-\beta_k \phi_{1,k}^{est})\ ,
\end{equation}
where $G_\beta$ is the gain of the ratio servo.  The scaling factor $\tilde{\nu}_1$ should be close to the frequency of clock 1, but it only modifies the gain of the ratio servo, so its accuracy is not critical.  Corrections are applied to the laser system itself via
\begin{equation}\label{eq-laser-freq-update}
c_{1,k+1} = c_{1,k} + \frac{G_1}{2 \pi T} \phi_{1,k}^{est} \ ,
\end{equation}
where $G_1$ is the gain of the clock 1 frequency servo. Here we have used the fact that the projection noise of clock 1 is much better than that of clock 2, so that only the phase measurements on clock 1 are relevant, but in principle, both can be used together to feedback on the laser.  In order to achieve enough feedback gain to overcome the long-time laser frequency drift, we often must include a second integrator for the laser frequency corrections in the Monte-Carlo model.  This is implemented by replacing Eq.~\ref{eq-laser-freq-update} above with
\begin{subequations}
\begin{align}
e_{1,k} &= \frac{G_1}{2 \pi T} \phi_{1,k}^{\rm est} \ , \\
c'_{1,k+1} &= c'_{1,k} + e_{1,k} \ , \\
c_{1,k+1} &= c_{1,k} + e_{1,k} + G'_1 c'_{1,k} \ ,
\end{align}
\end{subequations}
where $G_{1}^{\prime}$ is the gain of the second integrator.  A second integrator is not needed for the frequency ratio feedback for the noise we have considered.

%We estimate the phase $\phi_{j, k}^{\rm est}$ using the transition probability $p_{j,k}$, which is randomly chosen from a binomial distribution.  For clock 1, $\phi_{1, k}^{\rm est} = {\rm arcsin}(2 p_{1,k}-1)$ since the second $\pi/2$-pulse is shifted by a phase $-\pi/2$ with respect to the first.  The measurement result $\phi_{1,k}^{\rm est}$ is used to correct the phase of the clock 2 laser by applying a phase shift of $-\beta_k \phi_{1,k}^{\rm est}$ before its second Ramsey $\pi/2$ pulse, such that ${\phi_2^{\rm est} = \arcsin(2 p_{2,k} -1) + \beta \phi_1^{\rm est}}$, as depicted schematically in Fig.~\ref{figs-transition-probability}.  These measurement results are used together to perform feedback to the ratio according to
%\begin{equation}\label{eq-freq-ratio-update}
%\beta_{k+1} = \beta_k - \frac{G_\beta}{2\pi\tilde{\nu}_1 T} \left(\phi_{2,k}^{\rm est}-\beta_k \phi_{1,k}^{\rm est}\right)\ ,
%\end{equation}
%as well as a correction to the laser frequency according to
%\begin{equation}\label{eq-laser-freq-update}
%c_{1,k+1} = c_{1,k} + \frac{G_1}{2 \pi T} \phi_{1,k}^{\rm est} \ .
%\end{equation}
%Here, $G_1$ and $G_\beta$ are the gains of the clock 1 frequency servo and the ratio measurement servo, respectively, and the scaling factor $\tilde{\nu}_1$ represents the current best knowledge of the absolute frequency of clock 1, but its exact value is not critical to the operation of the feedback loop.

\section{\label{sec:GHZ}Measurements with Entangled States of Atoms}
\begin{figure}
\begin{center}
\includegraphics{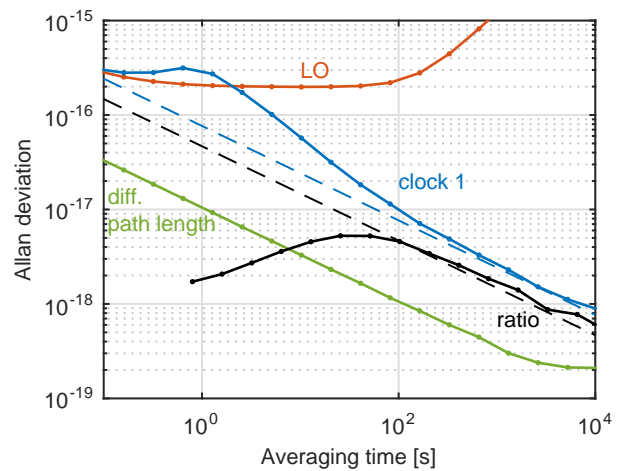}
\caption{\label{figs-monte-carlo-adaptive}Simulated stability of a comparison between a ytterbium optical lattice clock and an aluminum ion clock operating with five ions in a GHZ state.  The fractional clock 1 frequency stability (blue points) and the fractional ratio measurement stability (black points) are shown, along with the common-mode, unstabilized, laser frequency noise (red points) and the differential laser frequency noise (green points).  Clock 1 reaches the Dick effect stability limit (blue dashed line), while the ratio stability exceeds that, reaching the calculated projection noise limit for the aluminum ion clock (black dashed line).}
\end{center}
\end{figure}

 %A. Quantum Correlations and LO noise
The simulation results in Fig.~\ref{figs-noise-vs-ratio} extend to frequency ratios well beyond those available with the current generation of optical clocks.  However, a clock based on $N$ atoms prepared in a maximally entangled Greenberger-Horne-Zeilinger (GHZ) state operates effectively at a frequency $N$ times higher.  These states have been produced in the laboratory for small numbers of trapped ions up to $N = 14$ \cite{Monz2011}.  Previously, consideration of experimental noise sources including local oscillator noise has made the application of these quantum states for spectroscopy unrealistic for small numbers of atoms \cite{Huelga1997}.  Other quantum states and spectroscopy protocols have been proposed that retain some quantum advantage even in the presence of noise \cite{Andre2004, Rosenband2012, Mullan2014}, but none reaches the Heisenberg limit with realistic local oscillator noise.  Here, we show that frequency ratio measurements between two clocks with phase-locked local oscillators can take full advantage of the quantum-enhancement at the Heisenberg limit.

Consider the case where we replace the single-atom clock of Fig.~\ref{figs-noise-vs-ratio} with a clock based on $N$ atoms prepared in a GHZ state.  Such a clock has been shown, in principle, to provide Heisenberg-limited measurement variance~\cite{Bollinger1996},
\begin{equation}\label{eqn-heisenberg-limit}
\left(\frac{\Delta\nu}{\nu}\right)^2 = \frac1{(2\pi\nu)^2N^2T\tau} \ .
\end{equation}
For independent operation of a single clock, LO noise limits the probe time to ${T = \eta/(N\nu\sigma_L)}$, returning the measurement to the same projection noise limit as that for $N$ unentangled atoms [Eq.~(\ref{eq-proj-noise-1})]~\cite{Huelga1997}.  This has previously been confirmed numerically with a realistic laser noise spectrum~\cite{Rosenband2012}.  Note, however, that in our clock comparison protocol, the duration of the probe is limited not by laser noise but by the projection noise of clock 1, which may be orders of magnitude smaller.  Since a clock operating with atoms in a GHZ state evolves at an effective frequency $N \nu$, the performance of the comparison using our protocol can be determined from Fig.~\ref{figs-noise-vs-ratio} by substituting $\nu_2 \rightarrow N_2 \nu_2$.

\begin{figure}[!tb]
\begin{center}
\includegraphics{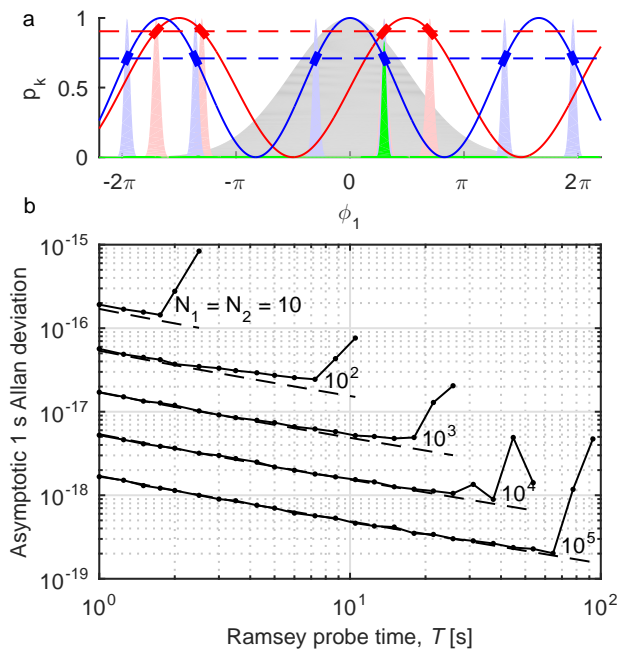}
\caption{\label{figs-monte-carlo-max-likelihood} Illustration of the maximum likelihood phase estimation algorithm. (a) The transition probabilities of clock 1 (red line) and clock 2 (blue line) are plotted as a function of the phase of clock 1 for zero frequency ratio error.  Example measurement outcomes for the two clocks are indicated by the horizontal dashed lines, with thick line segments indicating possible phase-inversion outcomes.  The prior distribution of laser phase values (gray), and clock 1 phase values (red) and clock 2 phase values (blue) based on the measurements are shown as shaded areas, with the product of the three shaded green.  (b) Simulated frequency ratio measurement stability of a strontium optical lattice clock with a ytterbium optical lattice clock using the maximum-likelihood protocol.  The projection noise limit is shown by dashed lines, and the simulation results are shown by solid lines.}
\end{center}
\end{figure}

%B. Model for entangled clock
We illustrate this with a detailed Monte Carlo simulation of a frequency comparison between a ytterbium optical lattice clock and an aluminum ion clock with five ions in a GHZ state.  For this simulation, in addition to the laser phase noise, we include differential phase noise due to path-length fluctuations between the two clocks, and we take into account the finite lifetime of the Al$^+$ $^3P_0$ state, dead time in both clocks, and the delay between the final $\pi/2$ pulses in the near-synchronous Ramsey experiments.  We assume that the Al$^+$ ions have been prepared perfectly in a GHZ state at the beginning of the Ramsey interval, and after the second Ramsey pulse, the parity of the atomic state is measured with unit fidelity~\cite{Bollinger1996}. In this case, during the Ramsey interval the atom state evolves as ${\ket{\psi(t)} = \left(\ket{\!\downarrow}^{\otimes N_2} + e^{-i\phi_2^{\prime}(t)}\ket{\!\uparrow}^{\otimes N_2}\right)/\sqrt{2}}$, where
\begin{equation}
\phi_2^{\prime}(t) = \int_0^t N_2 2\pi \left( \nu_2 - f_2(t^{\prime}) \right) {\rm d}t^{\prime} = N_2 \phi_2(t).
\end{equation}
The increase by a factor of $N_2$ in phase sensitivity must be reflected in the gain of the frequency ratio feedback such that Eq.~(\ref{eq-freq-ratio-update}) becomes,
\begin{equation}
\beta_{k+1} = \beta_{k} - \frac{G_\beta}{2 \pi \tilde{\nu}_1 T } \left(\frac{\phi_{2,k}^{\rm est}}{N_2}-\beta_k \phi_{1,k}^{\rm est}\right) \ .
\end{equation}
Since the GHZ state is also more sensitive to spontaneous decay, the lifetime of these states is modeled using $\tau_j \rightarrow \tau_j/N_j$.  The simulated ratio comparison stability shown in Fig.~\ref{figs-monte-carlo-adaptive} is found to be consistent with the Heisenberg limit for the Al$^+$ clock [Eq.~(\ref{eqn-heisenberg-limit})], with a small offset due to the finite lifetime of the $^3P_0$ state ($\tau = $ 20.6 s~\cite{Rosenband2007}).  This indicates that the ratio stability is reaching the limit imposed by the atomic coherence of Al$^+$.  The averaging period of 35 min to reach a statistical measurement uncertainty of $1\times10^{-18}$ is reduced by a factor of 25 from an asynchronous clock comparison with otherwise identical laser noise parameters.

\section{\label{sec:maximum-likelihood}Maximum-likelihood protocol}

%There are two possible outcomes of the maximum likelihood phase inversion algorithm in this scenario with non-negligible probability, indicated by the two green peaks.
So far, our discussion has focused on the use of one clock with low projection noise to reduce the projection noise of a second clock.  However, in a comparison between two clocks with low projection noise (e.g., two optical lattice clocks), it is possible to combine information from the two simultaneous phase measurements to extend the probe time of both.  Again, we consider a simultaneous Ramsey experiment on two clocks operating at different frequencies, but in this case the Ramsey probe duration $T$ extends beyond the limits imposed by LO noise for both clocks, and the phase estimate must be modified to accommodate clock phases outside the range $[-\pi/2, \pi/2]$.

For a given set of measurement outcomes $\{p_1, p_2\}$ of the two clocks, there are multiple sets $\{\phi_{1,n}^{\rm est},\phi_{2,m}^{\rm est}\}$ of the two clock phases indexed by $n$ and $m$, where ${\phi^{est}_{j,m} = m\pi + (-1)^m R^{-1}(p) - \theta_j}$.  Here, we have dropped the measurement index $k$ for convenience.  Note that the additional measurement phase for the second Ramsey $\pi/2$ pulse on clock 1 is always $\theta_1 = 0$ whereas, for clock 2, it is set to a random value $\theta_2$ for each probe in order to help avoid ambiguous phase inversion.  We calculate the statistical weight $W_{n,m}$ of each possible phase pair via a maximum likelihood analysis such that
\begin{equation}
W_{n,m} = \mathcal{N} \int_{-\infty}^{+\infty} d\phi_1 P_{1,n}(\phi_1) P_{2,m}(\phi_1) P_{L}(\phi_1),
\end{equation}
where
\begin{equation}
P_{L}(\phi_1) = \frac1{\phi_L \sqrt{2 \pi}}e^{-\frac{\phi_1^2}{2 \phi_L^2}}
\end{equation}
is the calibrated prior probability distribution for laser phase noise with standard deviation $\phi_L$, and
\begin{equation}
P_{j,n}(\phi_1) = \sqrt{\frac{N_j}{2\pi}} e^{-\frac12[\phi_j - \phi_{j,n}^{\rm est}(k)]^2 N_j}
\end{equation}
is the probability distribution centered at $\phi_{j,n}^{\rm est}$ based on the measurement result for clock $j$.  Here, $\mathcal{N}$ is a constant independent of $\phi_{1,n}^{\rm est}$ and $\phi_{2,m}^{\rm est}$ that can be determined by the normalization equality $\sum_{n,m} W_{n,m} = 1$.  These probability distributions are illustrated in Fig.~\ref{figs-monte-carlo-max-likelihood} (a).  The integral can be performed analytically giving
\begin{widetext}
\begin{equation}
W_{n,m} = \mathcal{M} \mathrm{exp}\left[-\frac{ \nu_1^2 (N_1 {\phi_{1,n}^{\rm est}(k)}^2 + N_2 {\phi_{2,m}^{\rm est}(k)}^2) + N_1 N_2 (\nu_2 \phi_{1,n}^{\rm est}(k) - \nu_1 \phi_{2,m}^{\rm est}(k))^2 \phi_L^2 }{ 2 \left[ \nu_1^2 + (N_1 \nu_1^2 + N_2 \nu_2^2) \phi_L^2 \right]  }\right] \ ,
\end{equation}
\end{widetext}
where $\mathcal{M}$ is a normalization constant.  Here, for the purposes of determining the proper feedback to both the laser and the frequency ratio, the atomic projection noise has been modeled as a Gaussian distribution of variance ${\left(\Delta \phi_j\right)^2 = 1/N_j}$ in phase for both clock 1 and clock 2.  This model is supported by the simulation, which uses them for calculating frequency corrections in the presence of realistic noise from atomic state projection and laser phase deviations.

The feedback corrections of the clock laser and the ratio are applied for all possible phase inversion outcomes, weighted by their normalized relative probability:
\begin{equation}
c_{1,k+1} = c_{1,k} + \frac{G_1}{2 \pi T} \sum_{n,m} W_{n,m} \phi_{1,n}^{\rm est}
\end{equation}
and
\begin{equation}
\beta_{k+1} = \beta_k - \frac{G_\beta}{2 \pi T \tilde{\nu}_1} \sum_{n,m} W_{n,m} \left( \phi_{2,m}^{\rm est} - \beta_{k} \phi_{1,n}^{\rm est} \right) \ .
\end{equation}
The summations in the above equations should in principle run over the range of all integers, but in practice can be truncated because $W_{n,m}$ is negligibly small for large enough $|n|$ or $|m|$.  The ranges ${n \in \{ -\textrm{ceil}(\frac{6 \phi_L}{\pi} - \frac{1}{2}) \ , \ \ldots \ , \ \textrm{ceil}(\frac{6 \phi_L}{\pi} - \frac{1}{2}) \}}$ and ${m \in \{ -\textrm{ceil}(\frac{6 \phi_L}{\pi} \frac{\nu_2}{\nu_1} - \frac{1}{2}) \ , \ \ldots \ , \ \textrm{ceil}(\frac{6 \phi_L}{\pi} \frac{\nu_2}{\nu_1} - \frac{1}{2}) \}}$, where ceil($\cdot$) denotes the ceiling function which rounds up to the next higher integer, cover the actual atomic phases with 6$\sigma$ confidence and are used in the Monte Carlo model.
%
%We apply a maximum likelihood analysis to determine the normalized relative probability $W_{n,m}$ of each possible phase pair (illustrated in Fig.~\ref{figs-monte-carlo-max-likelihood}(a)).  Feedback corrections of the clock laser and the ratio are applied for all possible phase inversion outcomes, weighted by their normalized relative probability:
%\begin{equation}
%c_{1,k+1} = c_{1,k} + \frac{G_1}{2 \pi T} \sum_{n,m} W_{n,m} \phi_{1,n}^{\rm est} \ ,
%\end{equation}
%and
%\begin{equation}
%\beta_{k+1} = \beta_{k} - \frac{G_\beta}{2 \pi \tilde{\nu}_1 T} \sum_{n,m} W_{n,m} \left( \phi_{2,m}^{\rm est} - \beta_{k} \phi_{1,n}^{\rm est} \right) \ .
%\end{equation}
%Using this maximum likelihood protocol, the probe duration can be extended to longer than the laser coherence time for both clocks, with a corresponding frequency ratio measurement stability given by the sum of the two projection noise contributions from Eq.~\ref{eq-proj-noise-1}.

The probe duration is limited by phase estimation errors caused by the projection noise of the two clocks.   Figure~\ref{figs-monte-carlo-max-likelihood} shows the asymptotic fractional ratio measurement stability for a comparison of a strontium optical lattice clock with a ytterbium optical lattice clock for different numbers of atoms.  With $N_1 = N_2 = 10$ atoms, the projection noise phase uncertainty is too large to allow unique inversion for phases outside the range $[-\pi/2, \pi/2]$, and the probe duration is limited to near the laser coherence limit at 1~s.  As the number of atoms is increased, degenerate inversion outcomes are less likely, and the probe time can be extended to longer than the laser coherence time, up to 30~s, for example, for $N_1 = N_2 = 10^4$.  The same clocks subject to identical laser noise but run asynchronously with a standard feedback routine give an asymptotic ($\tau = 1$~s) ratio stability of $4\times10^{-17}$, with the probe times limited to 1 and 1.2 s for clocks 1 and 2, respectively.  Thus, in this case, our protocol provides a reduction in the averaging time by a factor of 2000, with a factor of 100 improvement coming from the elimination of Dick effect noise and the remainder due to extending the Ramsey probe time.

\newcommand{\specialcell}[2][c]{\begin{tabular}[#1]{@{}l@{}}#2\end{tabular}}
\begin{table*}[ht!]
\begin{center}
\caption{\label{tab:MCparams} Parameters for numerical simulations.
}
\begin{tabular}{lll}
\hline\hline
				& Lattice-ion protocol (Fig.~3)		& Maximum likelihood protocol (Fig.~4) \\
\hline\hline
$N_1$				& 5000			& 10 to $10^5$ \\
$\nu_1$			& 518.296 THz		& 429.228 THz \\
$N_2$				& 5				& 10 to $10^5$ \\
$\nu_2$			& 1121.015 THz		& 518.296 THz \\
$T$				& 550 ms			& 1~s to 100~s \\
Dead time			& 250 ms			& 250 ms \\
\specialcell{Clock 1 measurement \\
\,\& feed forward time}	& 10 ms			& 0 \\
\specialcell{Clock 1 excited- \\
\, state lifetime} & 22.7 s \cite{Porsev2004} & $\infty$ \\
\specialcell{Clock 2 excited- \\
\, state lifetime} & 20.6 s \cite{Rosenband2007} & $\infty$ \\
\specialcell{Common-mode \\
\,laser frequency noise}
& $\sigma_y(\tau) = \left( 2 \times 10^{-16} \right) \left[ \left( \frac{0.3~\textrm{s}}{\tau} \right) + 1 + \left( \frac{\tau}{100~\textrm{s}} \right)^2 \right]^{1/2}$ \cite{Jiang2011}
& $\sigma_y(\tau) = \left( 1 \times 10^{-16} \right) \left[ \left( \frac{0.1~\textrm{s}}{\tau} \right) + 1 + \left( \frac{\tau}{1000~\textrm{s}} \right)^2 \right]^{1/2}$ \cite{Nicholson2012} \\
\specialcell{Differential \\
\,laser frequency noise}	& $\sigma_y(\tau) = \left( 1.4 \times 10^{-19} \right) \left[ \left( \frac{5.5 \times 10^3 \textrm{s}}{\tau} \right) + 1 \right]^{1/2}$ \cite{Foreman2007, Nicolodi2014} & 0 \\
\specialcell{Number of timesteps \\
\,simulated}			& $4 \times 10^5$		& $10^6$ (per point) \\
$G_1$				& 1				& 0.1 \\
$G'_1$			& 0.1			& 0.01 \\
$G_\beta$			& 0.03				& 0.05 \\
\hline\hline

\end{tabular}
\end{center}
\end{table*}

\section{Conclusion}

We have described protocols for frequency ratio measurements of optical clocks that use phase-locked LOs to reduce the projection noise by extending the probe time beyond the laser coherence time and eliminating noise due to the Dick effect.  We emphasize here that most of these improvements can be realized with laser systems at demonstrated levels of performance, which addresses an immediate issue for the present generation of optical clocks.  For example, the suppression of differential laser phase noise via active stabilization of optical paths (e.g. fiber noise cancellation) as well as laser stabilization via femtosecond combs is a standard technique in many labs.  One experimental challenge in implementing such a measurement is to integrate path-length stabilization seamlessly across the entire path from one atomic ensemble to the other.  In the case of Fig.~\ref{figs-monte-carlo-adaptive}, comparing an aluminum ion clock to a Yb lattice clock, relative phase stability must be maintained between the two experiments, spanning several wavelengths that connect the 578-nm Yb clock laser at the atomic ensemble to the 267-nm Al$^+$ clock laser where it probes the trapped ions.  While all components of this phase-stabilized frequency chain have been demonstrated, the full implementation will require careful consideration of the sources of differential noise in the system.  Similarly, while it remains challenging to produce GHZ states of trapped ions, a number of techniques have been demonstrated, with fidelities above 90\% for up to 6 ions in a linear chain~\cite{Monz2011}.  In Fig.~\ref{figs-monte-carlo-max-likelihood} we have ignored differential laser phase noise to explore the limits of phase inversion using a maximum likelihood analysis.  In order to realize a comparison with an Allan deviation below $1\times10^{-17}$ and an averaging time of 1 s, differential noise in the femtosecond comb frequency transfer as well as path length noise would have to be reduced below this level.  On the other hand, as we envision moving optical clocks out of the laboratory for applications such as relativistic geodesy, the ideas presented here significantly relax the requirements on laser coherence, enabling measurement stability at the current state-of-the-art with laser stability orders of magnitude worse, which might be attained in a robust package.

\begin{acknowledgements}
We thank T. Rosenband and D. Wineland for useful discussions related to this work.  We also thank K. Beloy and E. Knill for critical reading of the manuscript.  We acknowledge support from the Defense Advanced Research Projects Agency and the Office of Naval Research.  This work is a contribution of the U.S. government, not subject to U.S. copyright.
\end{acknowledgements}

\appendix
\section{\label{sec:appendix} Parameters for Numerical Model}
Here, we tabulate (Table~\ref{tab:MCparams})( parameters used for the Monte-Carlo simulations presented in Figs.~3 and 4 in the main text.  Where appropriate, experimental references for the sources of these parameters are given.

% We have investigated this limit using our Monte-Carlo model and found that a randomized measurement phase for the second $\pi/2$ pulse on clock 2 reduces the probability of phase inversion errors.
%Note that we have not included any spontaneous emission or differential laser frequency noise in these simulations in order to illustrate the potential of the protocol.  This will need improvement over the current state-of-the-art in order to achieve clock comparison stability as low as that shown in Fig.~4(b) for large atom numbers.

% Create the reference section using BibTeX:
%\clearpage
%\bibliography{C:/Al_svn/dleibran/Mendeley/library}
%

\end{document}